\documentclass[10pt]{iopart}
\usepackage{graphicx}
\usepackage{epstopdf}

\begin{document}

\title[]{Temperature and high fluence induced ripple rotation on Si(100) surface}

\author{Debasree Chowdhury, Biswarup Satpati \& Debabrata Ghose}

\address{Saha Institute of Nuclear Physics, 1/AF, Bidhannagar, Kol-64, India}
\ead{debasree.chowdhury@saha.ac.in, debu.sinp@gmail.com}
\vspace{10pt}
\begin{indented}
\item[]March 2016
\end{indented}

\begin{abstract}
Topography evolution of Si(100) surface due to oblique incidence low energy ion beam sputtering (IBS) is investigated. Experiments were carried out at different elevated temperatures from 20$^{\circ}$C to 450$^{\circ}$C and at each temperature, the ion fluence is systematically varied in a wide range from $\sim$ 1$\times$10$^{18}$cm$^{-2}$ to 1$\times$10$^{20}$cm$^{-2}$. The ion sputtered surface morphologies are characterized by atomic force microscopy and high-resolution cross-sectional transmission electron microscopy. At room temperature, the ion sputtered surfaces show periodic ripple nanopatterns and their wave-vector remains parallel to ion beam projection for entire fluence range. With increase of substrate temperature, these patterns tend to demolish and reduce into randomly ordered mound-like structures around 350$^{\circ}$C. Further rise in temperature above 400$^{\circ}$C leads, surprisingly, orthogonally rotated ripples beyond fluence 5$\times$10$^{19}$cm$^{-2}$. All the results are discussed combining the theoretical framework of linear, non-linear and recently developed mass redistribution continuum models of pattern formation by IBS. These results have technological importance regarding the control over ion induced pattern formation as well as it provides useful information for further progress in theoretical field.\\
\end{abstract}

%
%
%
%
%

\section{Introduction}

In recent days, ordered patterns in nanometer(nm) range on semiconductor surface has been pursuing immense technological applications in semiconductor devices, optoelectronics, spintronics etc \cite{kamins1997lithographic, bosi2010germanium}. Large scale fabrication of such nanopatterns with simultaneous control on their purity and uniformity demand a cost-effective and reproducible technique. In this respect, low energy ion beam sputtering (IBS) is known to be a promising method among other top-down and bottom-up approaches of nanostructurin \cite{biswas2012advances}. Depending on ion incidence angle, IBS can induce different nanoscale topographies, like one dimensional ripples (oblique incidence), two dimensional dots (normal incidence), cones or needle shaped structures (grazing incidence) etc. within very short time on a wide range of substrates in just one step i.e., simple exposure to ion beam. The patterns by IBS is thought to be generated by self-organization between curvature dependent ion erosion and diffusion based surface relaxations \cite{bradley1988theory, sigmund1969theory, sigmund1973mechanism}. Depending on ion incidence angle, the wave-vector of ripples can appear in either parallel or perpendicular direction of ion beam projection and those are generally addressed as (i) parallel (observed in the range of incidence angle 45$^{\circ}$- 65$^{\circ}$) and (ii) perpendicular mode (observed in the range of incidence angle 80$^{\circ}$-85$^{\circ}$) ripples respectively \cite{bradley1988theory}. Capability of IBS technique of nanopatterning fall short in application perspective due to upper surface amorphization of nanostructures. To get rid of this amorphization, sputtering at elevated substrate temperature can be a key factor. Thus, the studies on temperature induced morphological evolution during IBS is important from technological point of view. 

This study is also important for theoretical understanding as well because the complete understanding of pattern formation mechanism is still not clear. Recently, it has been shown by Madi et. al. that mass redistribution of surface adatoms plays dominant role in pattern formation mechanism rather than the curvature dependent ion erosion, especially in low energy ($<$1keV) regime of ion sputtering \cite{carter1996roughening, madi2011mass}. On the other hand, the surface relaxation mechanisms involved in pattern evolution process can be of different types like thermally induced surface diffusion \cite{bradley1988theory}, ion induced surface diffusion \cite{makeev2002morphology}, ion-enhanced viscous flow relaxation \cite{herring1950effect, mullins1959flattening, umbach2001spontaneous} etc. and among them which one would be dominating depends on experimental parameters substrate temperature and ion flux. Thus, the studies of surface morphological behaviour as function of different ion processing parameters like substrate temperature, ion sputtering time, ion energy, ion current density etc. helps to get a deeper understanding of pattern formation mechanism during IBS as well as it also helps to improve the control of IBS techniques in fabrication of well-ordered patterns. Unfortunately, due to experimental complexity or possibility of having impurity contamination, only a few studies are reported on the high fluence and temperature dependent nanopatterning on Si(100) surface by low energy ion beam sputtering \cite{chason2001dynamics, erlebacher1999spontaneous, gago2006temperature}. 

In this paper, we report the influence of temperature 23-450$^{\circ}$C on Si surfaces during 500 eV Ar$^+$ ion irradiation at incidence angle 65$^{\circ}$ for ion fluences 1$\times$10$^{18}$cm$^{-2}$ to 1$\times$10$^{20}$cm$^{-2}$. The experimental results show an orthogonal rotation of ripple nanopatterns at 400$^{\circ}$C beyond ion fluence 5$\times$10$^{19}$cm$^{-2}$. The morphology of modified surfaces were characterized by atomic force microscopy and their microstructural and chemical compositions were investigated by high-resolution transmission electron microscopy. All the results are discussed within the framework of linear and non-linear continuum models of pattern formation during ion beam sputtering including recently developed mass redistribution model. 

\section{Theories of pattern formation by IBS}

The first breakthrough in understanding the evolution of ripple morphology and its dependence on different ion beam  processing parameters like substrate temperature, ion energy and ion current density was achieved by Bradley \& Harper (BH) together and what they proposed is usually known as linear Bradley-Harper (BH) theory \cite{bradley1988theory}. Based on Sigmund's theory \cite{sigmund1969theory, sigmund1973mechanism}, Bradley \& Harper considers the surface patterning due to IBS is a balance of surface instability arising due to a self-organization process between curvature dependent ion erosion which results in surface roughening and surface relaxation processes which makes the surface smooth \cite{bradley1988theory}. Later on, to explain the coarsening in ripple wavelength and saturation of ripple amplitude at higher ion fluences, Cuerno and Barabasi had extended the BH theory where they introduced some non-linear terms which accounts the tilt dependency of sputtering yield \cite{makeev2002morphology}. Recently, Madi et. al based on Carter-Vishnyakov effect \cite{carter1996roughening} showed that the ballistic mass redistrubution originating from the diffusion of sputter induced near surface adatoms are dominating over curvature dependent ion erosion \cite{madi2011mass}. Thus, the temporal evolution of ripple height h at any position ($\vec{x}$, $\vec{y}$) on surface during IBS, can numerically be expressed  by equation \ref{eq_1}
\begin{eqnarray}
\label{eq_1}
\frac{\partial{h}}{\partial{t}}=&&-\nu_{0}+\nu'_0\frac{\partial{h}}{\partial{x}}+S_{x}\frac{\partial^2{h}}{\partial{x^2}}+S_{y}\frac{\partial^2{h}}{\partial{y^2}}+\frac{\lambda_x}{2}\left(\frac{\partial{h}}{\partial{x}}\right)^2\nonumber\\
&&+\frac{\lambda_y}{2}\left(\frac{\partial{h}}{\partial{y}}\right)^2-K\nabla^4h+\eta
\end{eqnarray} 
The $S_{x,y}$ term represents the sum of curvature dependent erosive and mass redistribution co-efficients i.e., $S_{x,y}=S_{x,y}^{eros.}+S_{x,y}^{redis.}$ in the parallel ($x$) and perpendicular direction ($y$) of the ion beam projection respectively \cite{bradley1988theory, madi2011mass}. $\nu_{0}$ is the rate of material erosion from surface at normal incidence and $\nu'_0$ contributes as the derivative of the angle dependent sputtering yield \cite{bradley1988theory}. $\lambda_{x,y}$ correspond to KPZ nonlinearities associated with tilt dependent local erosion yield in respective directions and $\eta$ is the noise term incorporating the stochastic nature of incident ions \cite{makeev2002morphology}. The term $K$ denotes the diffusion co-efficient for different surface relaxation processes which can be of different origin such as (i) Thermally activated surface diffusion (TSD) which is generally used to explain the high temperature and ion flux dependence of ripple wavelength during IBS. The concept of TSD comes from BH theory \cite{bradley1988theory}. Assuming temperature induced surface diffusion(TSD) is an isotropic phenomena, $K^{TSD}$ can expressed as $\frac{D_s\gamma\rho}{n^2k_B T}$, where the surface self diffusivity $D_s$ follows Arrhenius law with temperature as $D_s=D_0 exp(-\Delta E/k_B T)$, here $D_0$= const., $\Delta E$ is activation energy for surface diffusion, whereas $\gamma$, $\rho$ and $n$ designated as the surface tension co-efficient, the areal density of diffusing atoms and the target atomic density respectively. (ii) Ion induced surface diffusion $K^{ISD}$, which is basically the higher order expansion of erosion process and mathematically equivalent to surface diffusion means it just mimicking the surface diffusion and does not execute the real mass transport along the surface \cite{makeev2002morphology}. It depends on the parameters of ellipsoidal distribution of deposited ion energy, ion current density etc. and has no dependence on temperature. $K^{ISD}$ seems to be anisotropic in nature with respect to ion beam direction and its details are given in ref. \cite{makeev2002morphology}. (iii) Ion-enhanced viscous flow relaxation (IVF) which is driven by surface tension within thin ion damaged layer and depends on defect concentration generated within collision cascade \cite{herring1950effect, mullins1959flattening, umbach2001spontaneous}. $K^{IVF}$ is represented by $\frac{\gamma a^3}{\eta_s}$ where $a$ is ion energy penetration depth, $\gamma$ and $\eta_s$ presents surface tension and ion enhanced surface viscosity respectively. $K^{IVF}$ can be considered as isotropic and constant quantity as the associated parameters do not show any temperature dependence or anisotropic nature with respect to ion beam projection. This smoothing mechanism has also no dependency on ion current density.

During oblique incidence ion irradiation, regardless of the respective smoothing mechanism, the alignment of ripple wave-vector is determined by the larger absolute value between two co-efficients $-S_{x}$ and $-S_{y}$ \cite{bradley1988theory, madi2011mass} and the ripple wavelength $\Lambda$ is expressed by:
\begin{equation}
\label{Eq_2}
\Lambda_{x,y}=2\pi\sqrt{2K_{xx,yy}/{S_{x,y}}}
\end{equation}
In case of isotropic surface diffusion process, $K_{xx}$ and $K_{yy}$ both reduces into $K$ mentioned in eq. (1).
                                                                                                                          
\section{Experimental} 

Ultrasonically cleaned 1$\times$1 cm$^2$ p-type single crystal Si(100) wafers are irradiated by 500 eV Ar$^+$ ions at incidence angle 65$^{\circ}$ (with respect to surface normal) in a broad beam high current ion beam system (M/s Roth \& Rau Microsystems GmbH, Germany) \cite{chowdhury2015nanorippling}. An inductively coupled RF plasma discharge ion source having three-grid graphite optics system is employed to extract homogeneous ion beam of diameter 3 cm. During extraction of ion beam, the pressure within chamber was $\sim 10^{-4}$ mbar while usually the base pressure was $\sim 10^{-8}$ mbar.  The ion sputtering was performed at elevated temperature from 23-450$^{\circ}$C. The substrate temperature can be raised by using a radiation type heater placed at top of the chamber. For different temperatures, the ion fluence range is varied from $1\times10^{18}$ to $1\times10^{20}$ cm$^{-2}$ where current density was kept fixed around 1000 $\mu$A cm$^{-2}$. During ion bombardment a plasma bridge neutralizer (PBN) is used to avoid the divergence of ion beam. The ion sputtered modified surface was characterized using an ex-situ Veeco MMSPM NanoScope IV atomic force microscope (AFM) in ambient conditions, operating in tapping mode with Si cantilevers having nominal tip radius of 10 nm. Quantitative information regarding amplitude, periodicity, wavelength of surface features are extracted from the AFM images processed by Nanotec Electronica SL WSxM software (version 5.0 Develop 5.3) \cite{horcas2007wsxm}. The structural and chemical composition of ion sputtered surfaces were investigated by high-resolution transmission electron microscopy (HRTEM) carried out in a FEI, TECNAI G2 F30, S-TWIN microscope operating at 300 kV.

\section{Results and discussion}

The experimental results are sequenced as: Initially, the temperature induced morphological evolution of Si (100) surface for different ion fluence are explored. In next section, the experimental observations are disscused in context of different continuum models of pattern formation.

\subsection{Temperature and fluence dependent morphological behavior of Si surface}

\begin{figure}
\includegraphics[width=1.0\textwidth,natwidth=610,natheight=642]{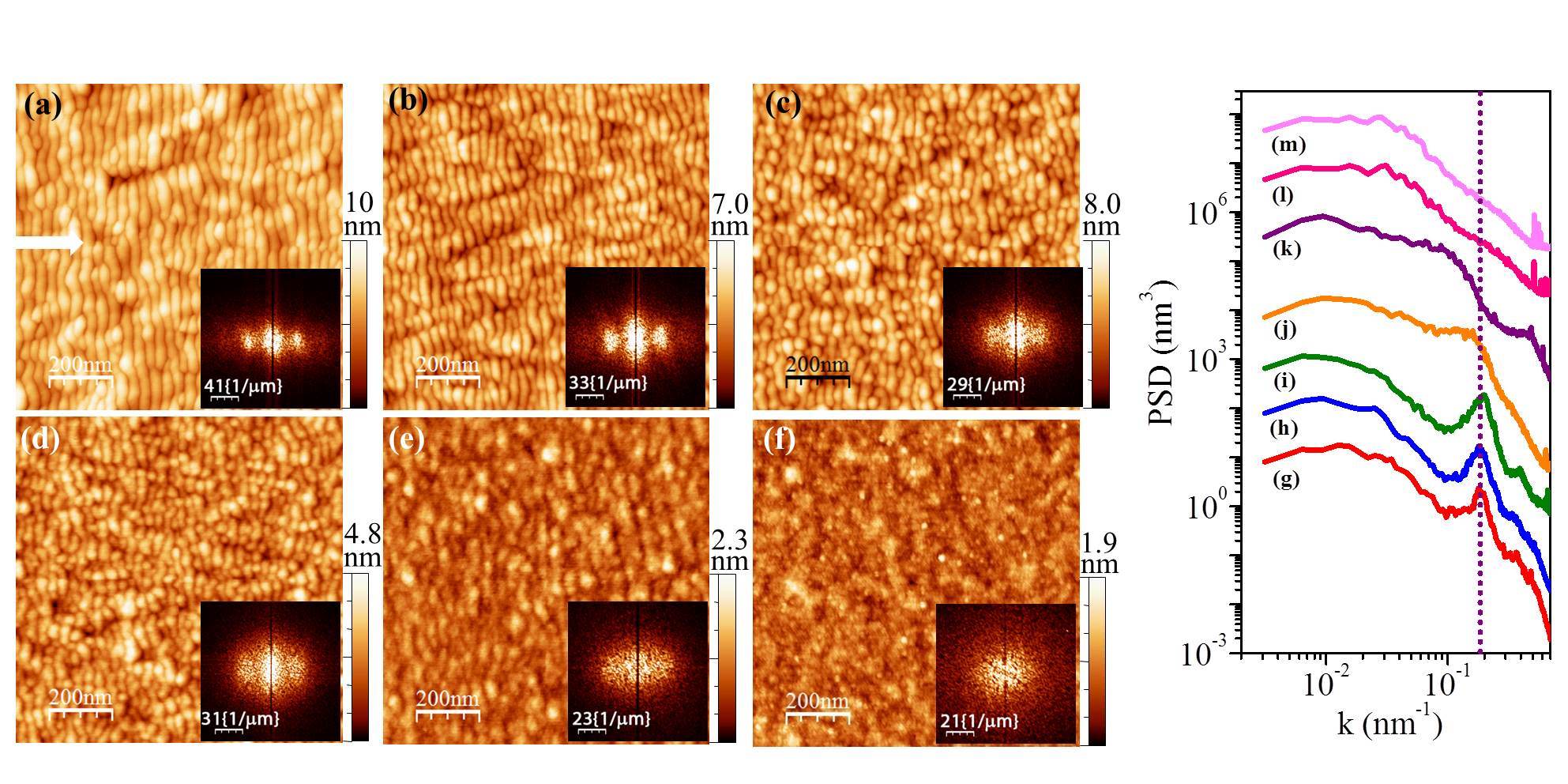}
\caption{AFM images of 500 eV Ar$^{+}$ sputtered Si surface mophologies at temperatures (a) 20$^{\circ}$C, (b) 100$^{\circ}$C, (c) 300$^{\circ}$C, (d) 350$^{\circ}$C, (e) 400$^{\circ}$C and (f) 450$^{\circ}$C for ion fluence $1 \times 10^{18}$ cm$^{-2}$. Inset shows FFT’s. Log-log plot of power spectrum density PSD(k, t) vs. spatial frequency k for sputtered Si surface at temperatures (g) 20$^{\circ}$C, (h) 100$^{\circ}$C, (i) 200$^{\circ}$C, (j) 300$^{\circ}$C, (k) 350$^{\circ}$C, (l) 400$^{\circ}$C and (m) 450$^{\circ}$C for ion fluence $1 \times 10^{18}$ cm$^{-2}$.}
\label{figure1}
\end{figure}

Fig. \ref{figure1} represents 1$\times$1 $\mu$m$^2$ AFM images of 500 eV Ar$^+$ ion sputtered Si(100) surfaces at different temperatures T = 20$^{\circ}$C, 200$^{\circ}$C, 300$^{\circ}$ C, 350$^{\circ}$C, 400$^{\circ}$C and 450$^{\circ}$C for low ion fluence $1\times10^{18}$ cm$^{-2}$. At room temperature $\sim$ 20$^{\circ}$C, the surface morphology exhibits well-ordered parallel mode ripple nanopatterns. Parallel mode means the wave-vector of these ripples are aligned parallel to ion beam projection. This can be confirmed by corresponding FFT image, where two 1$^{st}$ order bright spots are aligned symmetrically along the ion beam projection with respect to the central spot. As the temperature rises up to 200$^{\circ}$C, the ordering of ripple patterns remains as it is and both surface roughness and wavelength remain constant around 1.04 $\pm$ 0.08 nm 34.03 $\pm$ 0.60 nm (fig. \ref{figure4}c, \ref{figure4}d) respectively. Wavelength of ripples are determined from the first peak position of PSD functions shown in fig. \ref{figure1}(g-i) which have been extracted from the corresponding FFT images. The presence of sharp bright spots beside the center one in FFT image or the peak in PSD curves is the approval of periodicity on sample surface. From T $>$ 300$^{\circ}$C, the ripple amplitude begins to fall (fig. \ref{figure4}c) as well as ripples start to lose its periodicity (fig. \ref{figure1}). The mean surface roughness is found to decrease from 0.9 nm to 0.3 nm with rise of temperature from 300$^{\circ}$C to 450$^{\circ}$C. The loss of ripple ordering from 200$^{\circ}$C to 350$^{\circ}$C can be assured from the corresponding FFT images where the first order bright spots (Inset of fig. \ref{figure1}(c-d)) show gradual shrinkage of sharpness and, simultaneously, move towards closer to the central spot. The corresponding PSD curves (fig.\ref{figure1} (j-k)) also show relatively wider distribution of peak. The broad width means less ripple periodicity because FWHM of the peaks in PSD is inversely proportional to correlation length or surface periodicity. At 400$^{\circ}$C, the first bright spots finally merge with the central spot and FFT is left with an elliptical spot whose major axis lies along the parallel direction of ion beam projection. This reveals a poor degree of ripple ordering. As a consequence, corresponding PSD curve displays no peak (fig. \ref{figure1}i). Further increase of temperature up to 450$^{\circ}$C leads only a circular spot in FFT image. This indicates the presence of no preferred pattern orientation on surface. Consequently, the AFM image of fig. \ref{figure1}f corresponding to 450$^{\circ}$C also displays irregular modulations of very low amplitude surface heights. 

\begin{figure}
\includegraphics[width=1.0\textwidth,natwidth=610,natheight=642]{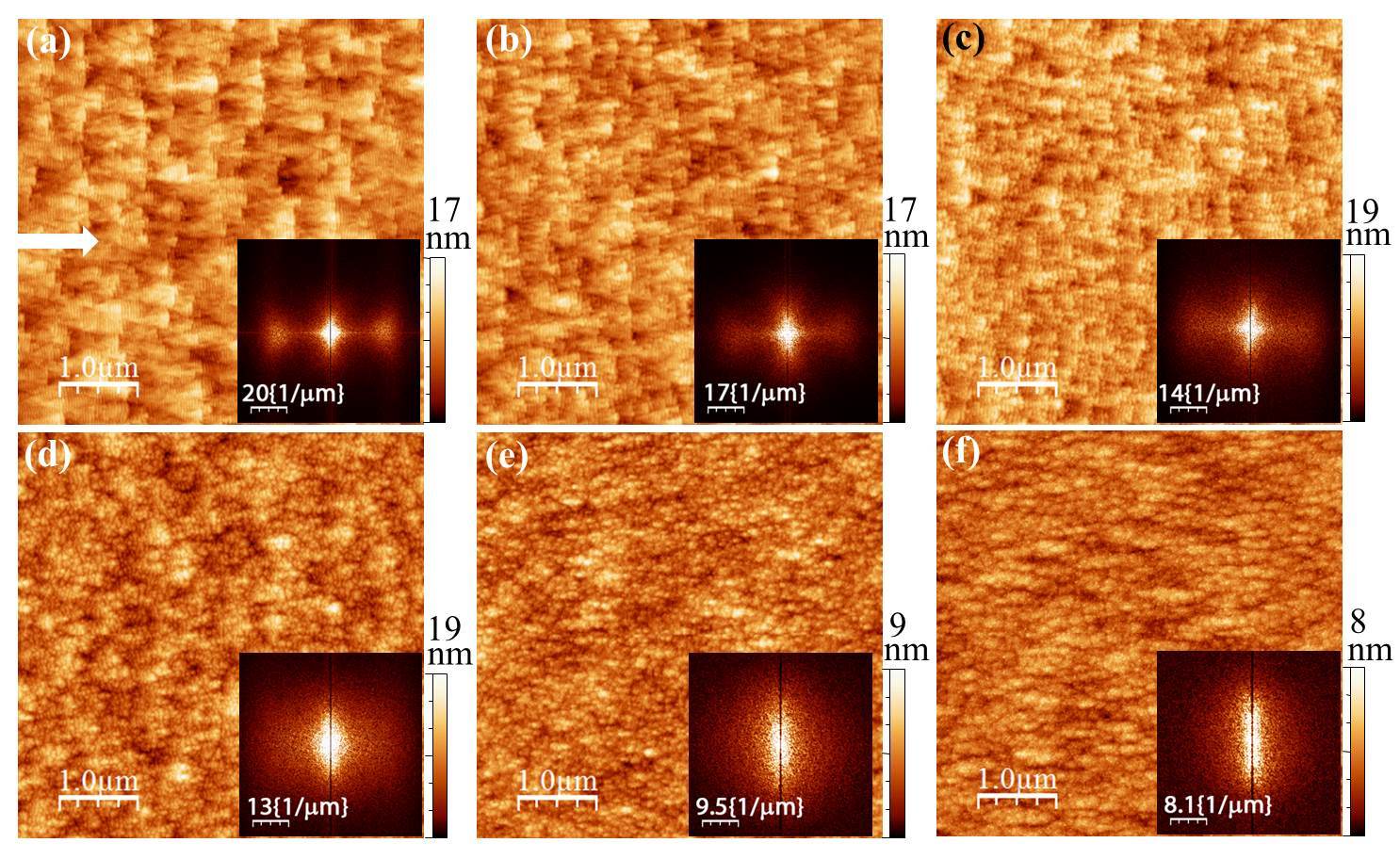}
\caption{5$\times$5 $\mu$m$^2$ AFM images show the evolution of Si surface morphologies for 500 eV Ar$^{+}$ ion sputtering at temperatures (a) 20$^{\circ}$C, (b) 200$^{\circ}$C, (c) 300$^{\circ}$C, (d) 350$^{\circ}$C, (e) 400$^{\circ}$ and (f) 450$^{\circ}$C for high ion fluence $1 \times 10^{20}$ cm$^{-2}$. Inset shows FFT’s. White arrow indicate the direction of ion beam projection.}
\label{figure2}
\end{figure}

In this section, the effect of high fluence on temperature induced Si surface morphologies are demonstrated. The experimental conditions were remain same as described in above paragraph, only ion fluence was varied up to 1$\times 10^{20}$ cm$^{-2}$. The surface morphologies at high fluence and at high temperatures are quite different to that observed for low fluence $1\times 10^{18}$ cm$^{-2}$. Some representative AFM images of temperature influenced surface pattens at high fluence $1\times 10^{20}$ cm$^{-2}$ are shown in fig. \ref{figure2}(a)–(f). The ion fluence dependent ripple morphological behaviours at room temperature are shown elsewhere \cite{chowdhury2013high}. Fig. \ref{figure2}(a) shows that the ripples, at room temperature, as a result of very high fluence $1\times 10^{20}$ cm$^{-2}$ retain their same orientation with respect to ion beam projection as in the low fluence regime (fig. \ref{figure1}). Also, a larger corrugation in wavelength is observed on surface together with the parallel mode ripple nanopatterns which develops beyond fluence $2\times 10^{19}$ cm$^{-2}$ (AFM images are not shown here) \cite{chowdhury2013high}. But as the temperature increases, due to high ion fluence $1\times 10^{20}$ cm$^{-2}$, the parallel mode ripples start to lose its ordering from 300$^{\circ}$C similar to low-fluence regime and reduce into randomly ordered mound-like structures around 350$^{\circ}$C. The larger corrugation in wavelength is also degraded with increase of substrate temperature and completely disintegrates at 350$^{\circ}$C. At T $\geq$ 400$^{\circ}$C, surprisingly, orthogonally rotated ripples are evolved where their wave-vector is oriented along perpendicular direction of ion beam projection (fig. \ref{figure2}e-f). These kind of ripples are generally addressed as perpendicular mode patterns. This can be assured by corresponding FFT image represented by fig. \ref{figure2}(e) which clearly displays the ellipsoidal distribution of the central spot along the perpendicular direction of the ion beam projection. Further rise in temperature (450$^{\circ}$C) leads to more pronounced perpendicular mode patterns. 

\begin{figure}
\includegraphics[width=1.0\textwidth,natwidth=610,natheight=642]{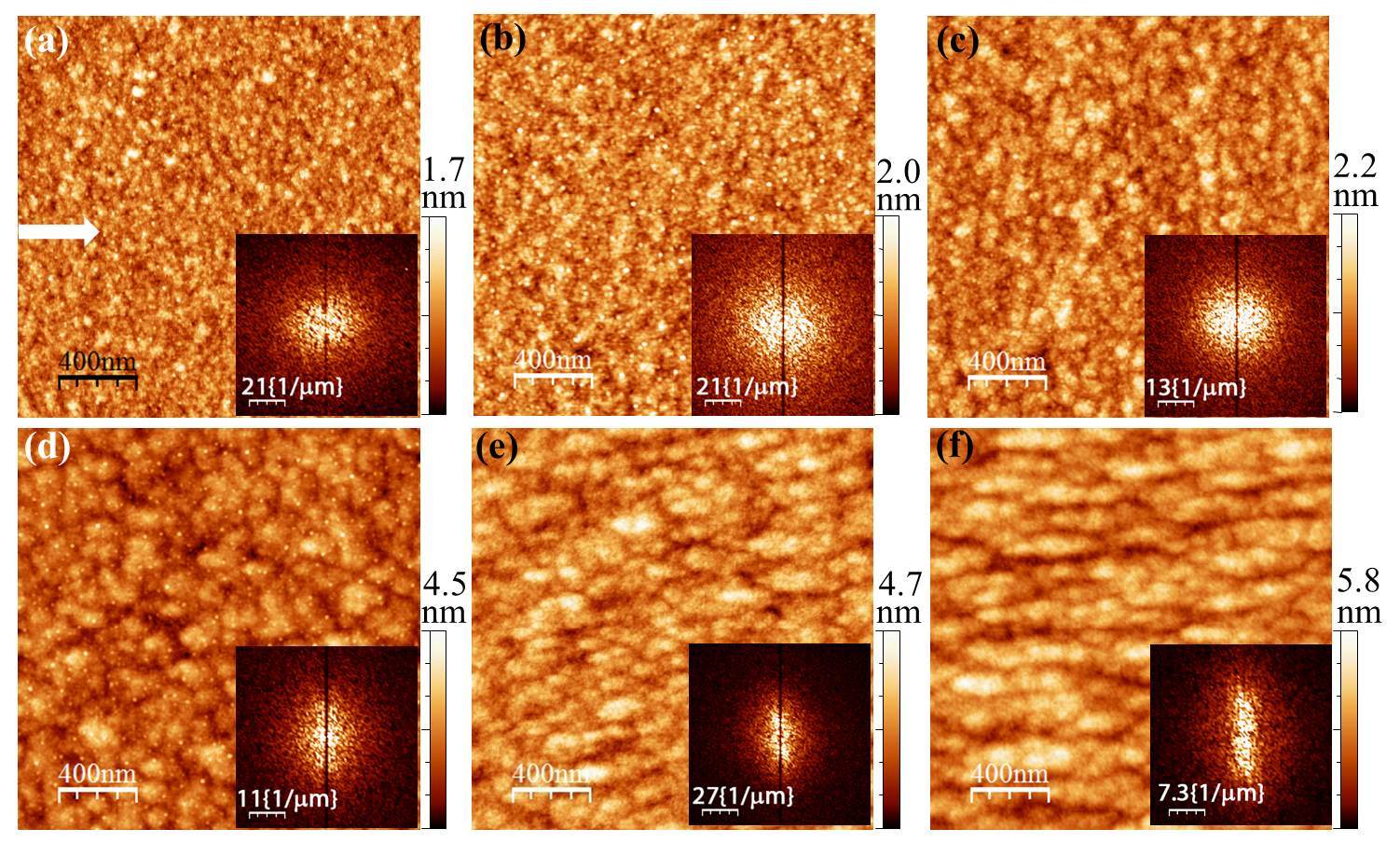}
\caption{ AFM images represent the evolution of perpendicular mode patterns at 450$^{\circ}$C for ion fluences (a) $1 \times 10^{18}$ cm$^{-2}$ (b) $5 \times 10^{18}$ cm$^{-2}$ (c) $1 \times 10^{19}$ cm$^{-2}$ (d) $2.3 \times 10^{19}$ cm$^{-2}$ (e) $5 \times 10^{19}$ cm$^{-2}$ and (f) $7 \times 10^{19}$ cm$^{-2}$. Inset shows corresponding FFT images. White arrow indicate the direction of ion beam projection.}
\label{figure3}
\end{figure}

\begin{figure}
\includegraphics[width=1.0\textwidth,natwidth=610,natheight=642]{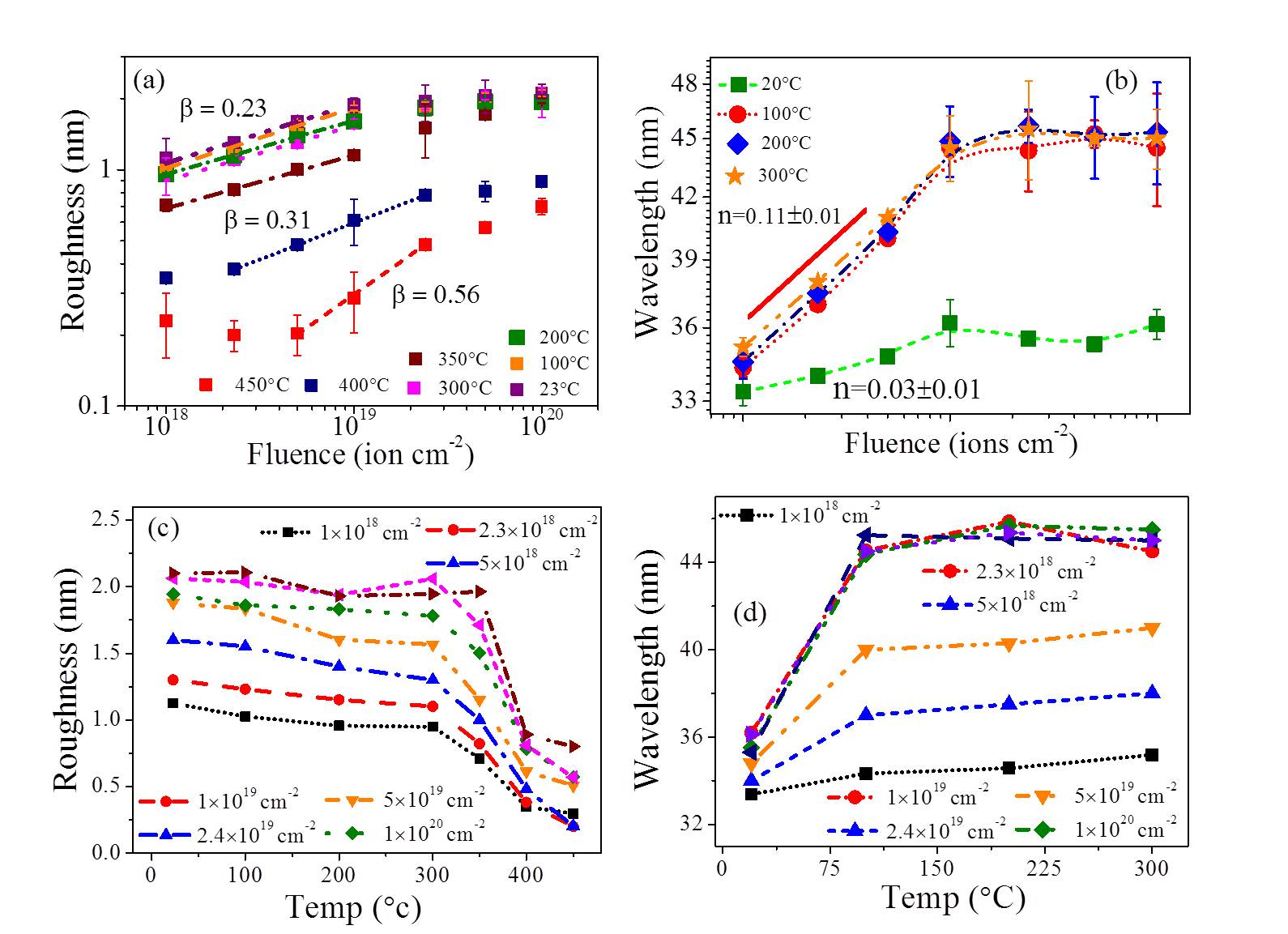}
\caption{ The variation of RMS roughness and ripple wavelength (a, b) with ion fluence by keeping the substrate temperature fixed and (a, b) with substrate temperature by keeping the fluence fixed.}
\label{figure4}
\end{figure}

The evolution of  perpendicular mode patterns at high temperature 450$^{\circ}$C as function of ion fluence are presented in fig. \ref{figure3}. The signature of perpendicular mode patterns is primarily observed for fluence $2\times 10^{19}$ cm$^{-2}$, although the pattern's alignment can not be visualized clearly from AFM image (fig. \ref{figure3}d) but realizable from the corresponding FFT image. These patterns become prominent with increase of ion fluence as can be seen from fig. \ref{figure3}(e) and fig. \ref{figure3}(f) respectively.

Based on fluence dependent changes in surface roughness $\omega$ up to 350$^{\circ}$C, the fluence range can be divide into two regimes: (1) first one is from $1\times 10^{18}$ cm$^{-2}$ to $1\times 10^{19}$ cm$^{-2}$ and (ii) second one is from $1\times 10^{19}$ cm$^{-2}$ to $1\times 10^{20}$ cm$^{-2}$ and are shown in fig \ref{figure4}(a). For the first regime, surface roughness is increased with growth exponent $\beta=0.23 \pm 0.01$ and remain almost constant in second regime. For temperature 400$^{\circ}$C and 450$^{\circ}$C, initially $\omega$ remains constant, then with fluence it follows a power law behaviour with exponent 0.31 $\pm$ 0.01 and 0.56 $\pm$ 0.02 respectively and get saturated at further high fluences. On the other hand, fig. \ref{figure4}(b) shows the ion fluence dependent behavior of ripple wavelength $\Lambda$ at different substrate temperatures from 20-450$^{\circ}$C where $\Lambda$ increases for ion fluence $1\times 10^{18}$ cm$^{-2}$ to $1\times 10^{19}$ cm$^{-2}$ and beyond that, it shows saturation. At room temperature, $\Lambda$ shows a very slow increase from 33 - 36 nm with coarsening exponent n = 0.03 $\pm$ 0.01 and at rest of the temperatures, it coarsen with exponent n= 0.11 $\pm$ 0.01. The temperature dependent surface morphological behaviours at different fluences are illustrated in Fig. \ref{figure4}(c) and (d). Irrespective of ion fluence, surface roughness is found to remain constant up to 350$^{\circ}$C and then falls at higher temperatures. On the other hand, for fluences $\geq$ $1 \times 10^{19}$ cm$^{-2}$, $\Lambda$ shows an initial increase with $T$ up to 100$^{\circ}$C and after that it becomes constant for higher $T$.

\begin{figure}
\includegraphics[width=1.0\textwidth,natwidth=610,natheight=642]{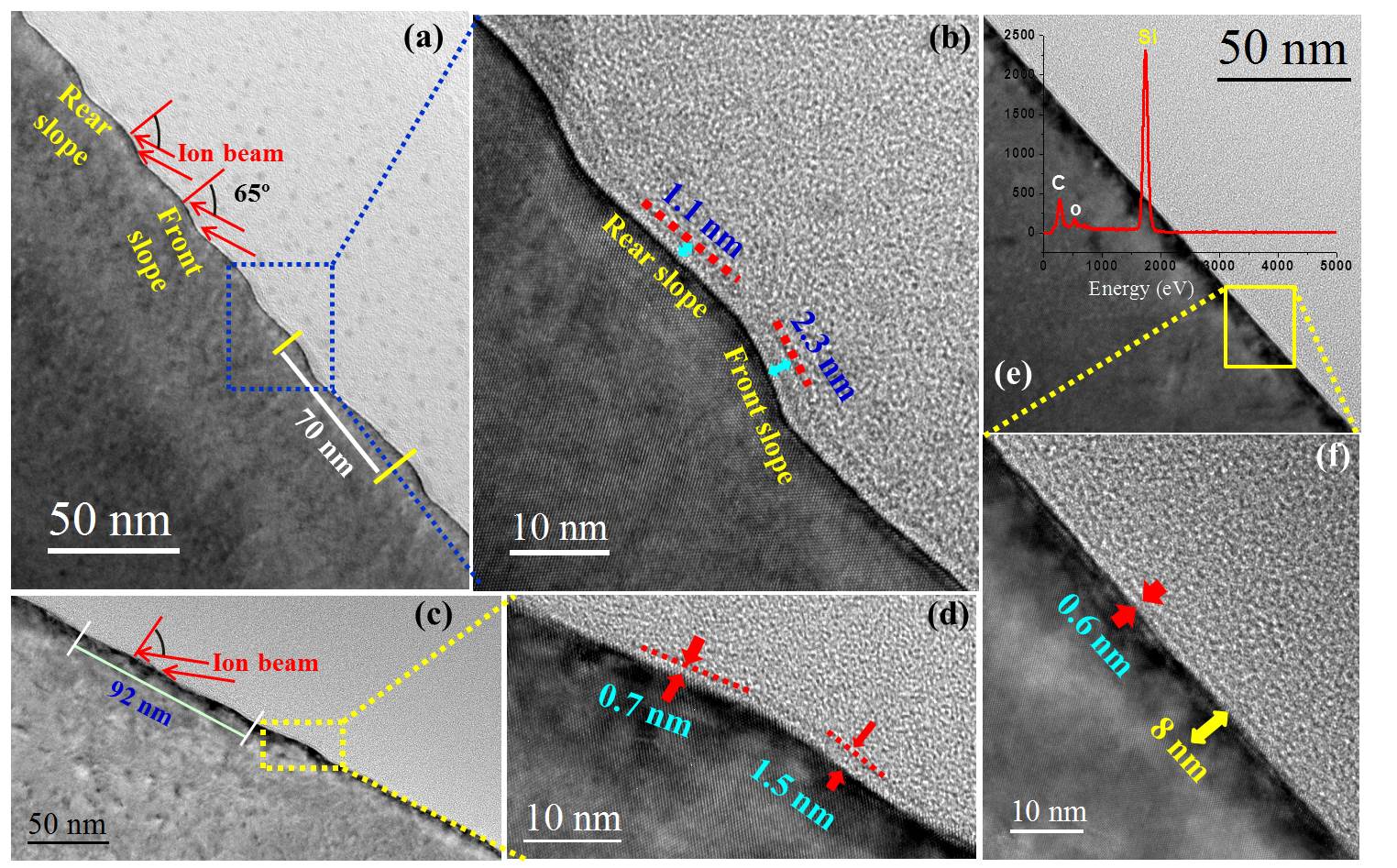}
\caption{ Microstructural characterization of 500 eV Ar$^+$ sputtered Si surface morphology at elevated temperatures for ion fluence 1$\times10^{20}$ cm$^{-2}$ and current density 1000 $\mu$A cm$^{-2}$. (a), (b), (c) and (d), (e) and (f) represent low magnification and high resolution cross-sectional TEM image of the surface morphology at 20$^{\circ}$C, 350$^{\circ}$C and 450$^{\circ}$C respectively.}
\label{figure5}
\end{figure}

In order to investigate the geometrical shape and crystallinity of temperature induced ion irradiated surface morphologies, cross-sectional transmission electron microscopy (TEM) is performed for the samples sputtered at highest ion fluence $1\times 10^{20}$ cm$^{-2}$ and at temperatures T = (i) 20$^{\circ}$C, (ii) 350$^{\circ}$C and (iii) 450$^{\circ}$C. The results are presented in fig. \ref{figure5}. The low magnified TEM image corresponding to fig. \ref{figure5}(a) shows ion sputtered Si surface at room temperature. The rippled surface profile does not exactly follow the sinusoidal nature but looks like sawtooth wave with average peak to peak separation 35 nm, one side of which facing the ion beam (front slope) is steeper than the other (rear slope). The high magnified TEM image (fig. \ref{figure5}b) explores a thin amorphous layer on crystalline Si with inhomogeneous distribution of thickness on front (2.3 nm) and rear slope (1.1 nm) of sawtooth ripples. The sawtooth shape of rippled surface profile and the inhomogeneity of amorphous layer may attribute to shadowing effect arising due to long time sputtering at grazing incidence (65$^{\circ}$). Due to shadowing effect, the front slope is eroded more and results steep edge with thick amorphous layer. The presence of amorphous layer on crystalline Si is quite expected due to irradiation at ion fluence $\sim 1\times 10^{20}$ cm$^{-2}$ which is sufficiently higher than the amorphization threshold \cite{drosd2010model}.

In comparison to room temperature, the TEM study of surface which is sputtered at 350$^{\circ}$C (Fig. \ref{figure5}c) shows quite low ordered and low amplitude height modulation in large lateral scale ($\sim$ 46 nm) which supports the morphologies revealed by the AFM image of fig. \ref{figure2}d). Simultaneously, the HRTEM image (fig. \ref{figure5}d) puts the evidence of narrow amorphous layer than at room temperature but it still shows non-uniformity in thickness at two slopes i.e., 1.5 nm and 0.7 nm for the front slope and rear slope respectively. At further higher temperature T = 450$^{\circ}$C (Fig. \ref{figure5}e), the surface becomes flat and covered with negligibly thin amorphous layer of 0.6 nm (Fig. \ref{figure5}f). Thus, the thickness of the surface amorphous layer is found to decrease with increase of substrate temperature for fixed high ion fluences. The outcomes of these TEM studies on temperature treated ion sputtered surfaces reveal one more interesting finding that the crystalline damage in near-surface region is found to increase with the increase of substrate temperature although the extent of damage is not uniform from the surface to inside bulk direction. In case of 450$^{\circ}$C, the damage is extended maximum up to 8 nm, while the penetration depth of 500 eV Ar$^+$ ions on Si surface incident at 65$^{\circ}$ is only of the order of 2 nm according to SDTrimSP V5.05 \cite{mutzke2011sdtrimsp}. This discrepancy can be explain in the following way: the above TEM studies already reveal the fact that the surface becomes more crystalline as the temperature rises. Thus, at higher temperature, it may possible that the ions can be channeled accidentally through some crystal sites and propagate upto more extent than its range. These cause the inhomogeneous damage in crystalline planes near the surface \cite{gemmell1974channeling}. 

In order to explore whether the temperature induced surface morphologies are not induced by impurity contaminants, high-angle-annular-dark field (HAADF) analysis, which essentially gives Z-contrast imaging, is performed. The spectrum of elemental mapping by energy dispersive x-ray spectroscopy (EDX) over a domain (marked in Fig. \ref{figure5}e) of ion sputtered surface topography at 450$^{\circ}$C shows strong Si lines with weak signals of C and O at the background shown in fig. \ref{figure5}(e) and no evidence of any metal species are found. The presence of adsorbed species C and O are probably come due to exposure of the bombarded sample into the air.

\subsection{Theoretical discussions on surface morphological behaviours}

\begin{figure}
\includegraphics[width=0.6\textwidth,natwidth=610,natheight=642]{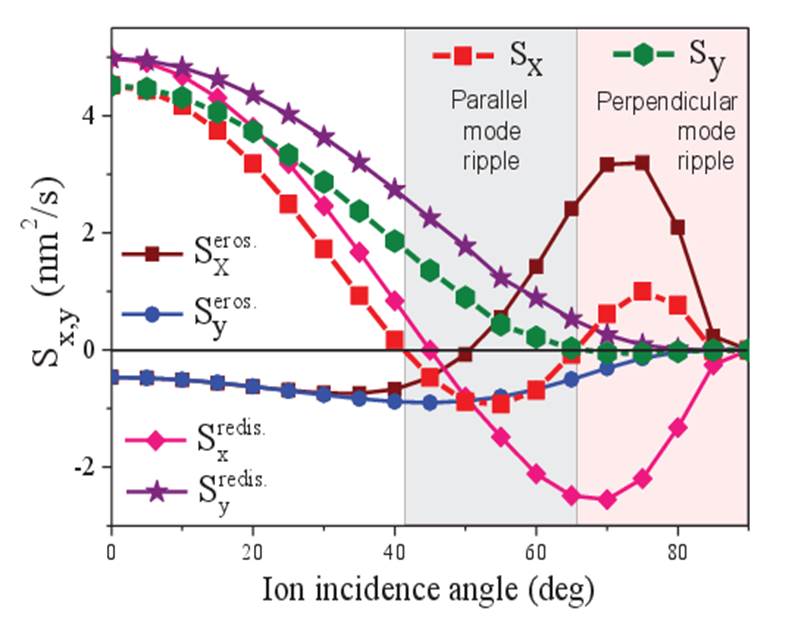}
\caption{ Calculated values of $S_{x,y}$, $S^{eros.}_{x,y}$ and $S^{redis.}_{x,y}$ vs. ion incidence angle.}
\label{figure6}
\end{figure}

Some of the experimental results described above can be fitted well to the prediction of theoretical framework based on BH theory \cite{bradley1988theory}, whereas, some other results do not. BH theory is already discussed in sec. 2, according to which pattern formation during sputtering is a complex interplay between curvature dependent ion erosion and smoothening via surface diffusion. Recently, Madi et. al. had shown that mass re-distribution of surface adatom plays dominant role in pattern formation mechanism rather than ion erosion. The larger absolute value between ion erosion $-S_{x,y}^{eros.}$ and mass-redistribution $-S_{x,y}^{redis.}$ co-efficients determine the formation of nanopattern and its orientation with respect to ion beam projection, where $x$ and $y$ signifies the same direction as described earlier. $S_{x,y}^{eros.}$ and $S_{x,y}^{redis.}$ can be calculated by ellipsoidal ion energy distribution parameters, average depth $a$ = 1.38 nm, lateral and longitudinal widths of energy distribution $\sigma$ = 1.19 nm and $\mu$ = 0.87 nm respectively and their angle dependent variations are shown in fig. \ref{figure6}. The parameters $a$, $\sigma$ and $\mu$ for 500 eV Ar$^+$ ion irradiation on Si surface are extracted using the code SDTrimSP V5.05 \cite{mutzke2011sdtrimsp}.
Fig. \ref{figure6} shows that $S_{x}^{redis.}$ exhibits larger negative value for $\alpha_{ion}$ = 65$^{\circ}$. This predicts the evolution of anisotropic patterns with its wave-vector along x direction which matches the experimental results and the pattern formation is dominated by the mass redistribution of near-surface adatoms rather than the curvature dependent erosion. 

On the other hand, the surface smoothening mechanism during sputter induced pattern evolution process via three different diffusion processes is also quite complex. Details of different smoothening mechanism is also discussed in sec 2. At room temperature, thermally induced surface diffusion ($K^{TSD}=4\times10^{-8}$ nm$^4$ s$^{-1}$ contributes negligibly small, whereas, ion beam induced surface diffusion $K_{xx}^{IESD}$ gives the ripple wavelength (using eq. 2) one order less than the experimental value. By considering ion induced viscous flow $K^{IVF}$ as smoothening mechanism, the estimated ripple wavelength for parameters $\eta_Si$=4.31 nm$^{4}$ s$^{-1}$, $\gamma$=6.74 eV nm$^{-2}$ and N=49.77 atom.nm$^{-4}$ is found to tally with the experimentally obtained value. Thus, $K^{IVF}$ can be concluded as dominant surface relaxation mechanism during IBS at room temperature. For higher temperatures, K$^{TSD}$ increases following Arrhenius law but its contribution in pattern formation up to 300$^{\circ}$C ($\sim$ 3.6 nm$^4$ s${^{-1}}$) is too small. Further rise of temperature obviously makes TSD more strong ($\sim$ 187.6 nm$^4$ s$^{-1}$) but somehow the experimental results do not follow the predicted BH behavior which is the increase of ripple wavelength with temperature \cite{bradley1988theory}. In contrast, the experimental results show the degradation of parallel mode ripples as temperature is increased. Other smoothening mechanism, \textit{i.e.,} K$^{ISD}$ exhibits no dependency on temperature. Thus, it may possible that IVF is playing the smoothening mechanism during temperature induced pattern formation. But, as the exact temperature dependence of IVF mediated parameters like surface tension and viscosity are still lacking information, the mechanism of smoothening by IVF at high temperature is not clear. The fall of surface roughness and disappearance of parallel mode patterns at higher temperatures may driven by some other phenomena rather than the conventional diffusion process discussed above. One of the possibility is the re-crystallization of ion beam induced amorphization due to high temperature \cite{brown2005temperature}. During reconstruction of amorphized layer into its crystalline phase, strain fields are developed among crystal planes. The compressive or tensile strain can either increase or decrease the surface diffusion depending on whether the adatom-induced surface stress is under tension or compression \cite{park1999dynamics}. Gago \textit{et al.} \cite{gago2006temperature} had confirm the presence of compressive strain in near surface region through GID measurements.The TEM studies of ion irradiated Si(100) surfaces at different elevated temperatures (Fig. \ref{figure5}) show increase of defects in near surface crystalline planes with increase of temperature which may arise due to presence of these strain effects. 

On the other hand, the temperature dependent orthogonal rotation of ripple patterns on Si(100) surface is quite surprising. The possibility of getting rotated mode ripples is predicted by non-linear extension of BH theory. The negative sign of the product of non-linear terms $\lambda_{x}$ and $\lambda_{y}$ predicts the rotation of ripple by an angle $\theta_{c}= \tan^{-1} (\lambda_{x,y}/\lambda_{y,x})$ after a prolonged sputtering time $t_c\propto\frac{K}{(S_{x,y})^{2}} \ln \frac{S_{x,y}}{\lambda_{x,y}}$ \cite{park1999dynamics} at room temperature. In present case, the experimental results show ripple rotation about 90$^{\circ}$ after fluence $2 \times 10^{19}$ cm$^{-2}$ ($\sim$ 2 hr.) for substrate temperature 450$^{\circ}$C. Here, the ripple rotation is caused by both high temperature and high sputtering time. This behavior is quite surprising and can not be explained by any existing continuum model of pattern formation till now. Temperature dependent evolution of perpendicular mode ripples due to high ion fluence may attributed to the temperature induced diffusion of surface adatoms. As temperature increases, the adatoms mobility increase and they can move freely over the surface. In this scenario, the adatoms can be easily drifted by any net force acting on them. During oblique incidence ion sputtering, the transfer of momentum from incident ions to surface atoms parallel to sample surface exerts a net amount of force on surface. As a result, the substrate atoms as well as adatoms diffuse along the direction of impinging force and, due to long time of ion bombardment, the diffused adatoms may generate patterns aligned parallel to ion beam projection means the pattern's wave-vector is lying along perpendicular direction of ion beam projection. This phenomena may cause the high fluence induced perpendicular mode ripples at high temperature. On the other hand, the temperature and fluence induced orthogonal rotation had reported by only one group, Erlebacher \textit{et al.} where they conclude that the surface diffusion via surface vacancies dominates the pattern formation mechanism \cite{brown2005transient}. Therefore, it may possible for our case also that instead of dimers, which are identified as the likely diffusing species on Si surface \cite{shu2001simple}, the vacancies are playing major role in surface diffusion at high temperature. Moreover, the evolution of perpendicular mode ripple patterns are not as strong in our studies as observed in ref. \cite{brown2005transient}. This may attributed to lower temperature range used in our experiments compared to the reported value.  

\section{Conclusion}

The present work describes a systematic study on temperature induced morphological evolution of ion sputtered ripple surface for low to high fluence regime. For low fluence regime, self-organized parallel mode ripple is found to flatten with temperature. But surprisingly, ripples are orthogonally rotated at higher temperatures ($\sim 450^{\circ}$ C) after a certain time of sputtering. 350$^{\circ}$ C is identified as transition region of pattern orientation for entire fluence range. The EDAX study confirms that the surface morphologies are not induced by impurity or any other metal contamination. Apart from morphological changes, TEM study reveals that the crystalline defects in near-surface region is increased with increase of elevated substrate temperature. The experimental results are discussed in the framework of existing continuum models dealing with the morphology evolution of ion sputtered surfaces. The ion beam induced pattern formation is found to be dominant by mass-redistribution of surface adatoms rather than curvature dependent ion erosion. On the other hand, ion induced viscous flow as a dominant smoothing mechanism during pattern evolution. Some of the results presented here are surprising and demand further theoretical progress for clear understanding of ion induced pattern formation mechanism which in other way helps to improve the control of IBS technique on surface nanopatterning. 

\section{References}

\section{\label{sec:level1}References}

\end{document}